\documentclass[draft,final ]{aipproc}

\usepackage{amsmath,amssymb,bm}
\usepackage{graphicx}
\usepackage{latexsym}

\layoutstyle{8x11double}

\def\nn{\nonumber}

\let\p=\partial
\def\tk{\tilde{k}}
\def\be{\begin{equation}}
\def\ee{\end{equation}}
\def\bea{\begin{eqnarray}}
\def\eea{\end{eqnarray}}
\def\ba{\begin{array}}
\def\ea{\end{array}}

\def\vep{{\varepsilon}}

\begin{document}

\title{Relativistic magnetotransport in graphene}


\author{Markus M\"uller}{
address={Department of Physics, Harvard University, Cambridge MA 02138, USA}
}
\author{Lars Fritz}{
address={Department of Physics, Harvard University, Cambridge MA 02138, USA}
}
\author{Subir Sachdev}{
address={Department of Physics, Harvard University, Cambridge MA 02138, USA}
}

\author{J\"org Schmalian}{
address={Ames Laboratory and Department of Physics and Astronomy, Iowa State
University, Ames, IA 50011, USA}
}

\begin{abstract}
We study the thermal and electric transport of a fluid of interacting Dirac fermions as they arise in single-layer graphene. We include Coulomb interactions, a dilute density of charged impurities and the presence of a magnetic field to describe both the static and the low frequency response as a function of temperature $T$ and chemical potential $\mu$.
In the critical regime $\mu\lesssim T$ where both bands above and below the Dirac point contribute to transport we find pronounced deviations from Fermi liquid behavior, universal, collision-dominated values for transport coefficients and a cyclotron resonance of collective nature. In the collision-dominated high temperature regime the linear thermoelectric transport coefficients are shown to obey the constraints of relativistic magnetohydrodynamics which we derive microscopically from Boltzmann theory. The latter also allows us to describe the crossover to disorder-dominated Fermi liquid behavior at large doping and low temperatures, as well as the crossover to the ballistic regime at high fields.
\end{abstract}

\maketitle

\section{Introduction}
The quasiparticles in single layered graphene obey the massless Dirac equation~\cite{Semenoff84,Haldane88,Zhou06}. At finite temperature and moderate doping, they form an interacting electron-hole plasma of ultra-relativistic fermions whose transport properties differ significantly from a conventional Fermi liquid.
Indeed, it has been argued that intrinsic graphene at $T=0$ realizes a system at a quantum critical point which controls the transport properties in a window of temperature and chemical potential given by $T\gtrsim \mu$ ~\cite{joerg,qcgraphene}. This criticality is reflected
in the inelastic scattering rate being proportional to $\alpha^2 T$ (we set $k_B=\hbar =1$ throughout), {\it i. e. }, solely depending upon temperature. Here, $\alpha$ is the fine structure constant characterizing the strength of Coulomb interactions, $\alpha=e^2/\kappa v$, where $\kappa$ is the dielectric constant of the adjacent medium and $v$ is the Fermi velocity of the linearly dispersing quasiparticles.

In this paper we show on the basis of a microscopic Boltzmann equation approach that at low frequencies and high enough temperatures graphene behaves like a plasma of charged particles and antiparticles obeying relativistic magnetohydrodynamics~\cite{cyclotron}. The temperature has to be high enough
such that the inelastic scattering rate $\tau^{-1}_{\rm ee}$ is bigger than the driving a.c. frequency $\omega$, and the scattering rates due to impurities, $\tau^{-1}_{\rm imp}$, or weak magnetic fields, $\tau^{-1}_B=v^2 e B/cT$.
All the frequency dependent thermo-electric conductivities turn out to be completely fixed by thermodynamic quantities, the covariant equations of motion and a single transport coefficient $\sigma_Q(\mu/T)$ which assumes a universal, entirely collision-dominated value in the critical regime $\mu/T \ll 1$ of a clean system~\cite{qcgraphene,kashuba}.

A logarithmic divergence in the collinear scattering amplitude, occurring generally in two-dimensional systems, allows us to obtain the transport coefficients to leading order in an expansion in the logarithm of the relevant infra-red cutoff, which in graphene turns out to be $1/\log(\alpha)$. This approach also allows for an elegant description of the crossover from the critical regime $\mu< T$ to the Fermi liquid regime $\mu\gg T$, establishing the connection with existing results~\cite{Nomura,DasSarmaGalitski}. To leading logarithmic order the collinear divergence results in an effective equilibration among quasiparticles moving in the same direction. This reduces the complexity of solving the Boltzmann integral equation for the distribution function to the determination of at most three numbers that characterize the angle-dependent effective equilibria of collinearly moving particles.

Very similar transport characteristics as in the critical region of graphene is expected
in other relativistic systems as they often emerge close to quantum critical points with a dynamical exponent $z=1$, such as the superfluid-insulator transition~\cite{bhaseen,nernst} in a Bose-Hubbard model. Boltzmann transport theory has first been applied to such critical systems by Damle and Sachdev~\cite{damless,ssqhe}, with the important difference that there only short-range interactions between the linearly dispersing bosonic excitations were considered. Bhaseen et al.~\cite{bhaseen} have recently extended this approach, discussing interesting effects in the presence of a magnetic field.
With minor modifications, the present general analysis applies to such systems, too.
A detailed analysis of Boltzmann magnetotransport of relativistic quasiparticles in a magnetic field was given in Ref.~\cite{MHDgraphene}.

\section{Relativistic hydrodynamics from Boltzmann approach}
If the Coulomb interactions are not too strong, $\alpha \ll 1$, a description of the system in terms of quasiparticles is possible. In graphene, this is certainly the case if the sample is adjacent to a dielectric with $\kappa\gg 1$, but also holds generally at low enough temperatures, since the coupling $\alpha$ is marginally irrelevant and flows logarithmically to zero as $T\to 0$, see~\cite{guinea,son}. This allows us to analyze transport within a quasiclassical Boltzmann approach, solving for the distribution function $f_{{\bf k},\lambda}({\bf r},t)$ in the presence of  driving fields which vary slowly in space and time.

The low energy excitations belong to two valleys around two inequivalent Dirac points ${\bf K,K}'$. Within each valley the fermionic quasiparticles form two bands with linear dispersion $\varepsilon_{\bf k,\lambda}=\lambda v k$, where $\lambda=\pm 1$ refers to the band of electrons and holes of the undoped material, respectively.

In equilibrium the distribution function is
\begin{equation}
f^0_{k,\lambda}=\frac{1}{\exp[(\varepsilon_{\bf k,\lambda}-\mu)/T]+1}.
\label{feq}
\end{equation}
In the presence of an electrical field ${\bf{E}}(t)$, a temperature gradient $\nabla T$, and a perpendicular magnetic field ${\bf{B}}=B{\bf e}_z$, the distribution function obeys the Boltzmann equation:
\bea
\left( \frac{\partial }{\partial t}+ e \left( \mathbf{E}+ {\bf{v}}_{{\bf{k}},\lambda} \times \frac{\mathbf{B}}{c}\right)\cdot \frac{\partial }{
\partial \mathbf{k}} + {\bf{v}}_{{\bf{k}},\lambda}\cdot \frac{\partial }{
\partial \mathbf{r}}\right) f_{{\bf k},\lambda}(\mathbf{r},t)\nn\\
=-{\cal I}_{\rm coll}[\lambda,\mathbf{r},\mathbf{k},t\,|\{f\}].  \label{trans0}
\eea
Here ${\cal I}_{\rm coll}[\lambda,\mathbf{r},\mathbf{k},t\,|\{f\}]$ denotes the collision integral due to Coulomb interactions and impurity scattering, and ${\bf{v}}_{{\bf{k}},\lambda}\equiv v\hat{\bf v}_{{\bf k},\lambda}=\nabla_{\bf k}\varepsilon_{{\bf{k}},\lambda}$, 
 denotes the quasiparticle velocity.
Assuming small deviations from equilibrium the above can be rewritten as
\begin{eqnarray}
\label{Boltzmann}
\left(\partial_t + \frac{e B}{c} \left({\bf{v}}_{{\bf{k}},\lambda}\times {\bf{e}}_z\right)\cdot \frac{\partial}{\partial {\bf{k}}} \right)f_{{\bf k},\lambda}(\mathbf{r},t)+{\cal I}_{\rm coll}[\{f\}]\\
 = T \left[ \hat{\bf{v}}_{\bf{k},\lambda} \cdot {\bf F}^E \phi^{E}({\bf k},\lambda)+  \hat{\bf{v}}_{\bf{k},\lambda} \cdot {\bf F}^T \phi^{T}({\bf k},\lambda)\right],\nn
\end{eqnarray}
where the RHS describes the driving terms due to small electric fields, and a slowly varying temperature, as characterized by the dimensionless vectors ${\bf F}^E=v e {\bf{E}}/T^2$ and ${\bf F}^T=-v {\bf \nabla}T/T^2$. Further, we have defined
\bea
\label{phis}
\left ( \begin{array} {c}\phi^{E}({\bf k},\lambda) \\ \phi^{T}({\bf k},\lambda) \end{array} \right) =f^0_{k,\lambda}(1-f^0_{k,\lambda})\cdot \left (  \begin{array}{c} 1\\ (\varepsilon_{\bf k,\lambda}- \mu)/T \end{array} \right).
\eea


In linear response we can analyze each driving term separately. For the corresponding driving force we will simply write ${\bf F}$ below. In linear approximation we may parameterize the deviations from local equilibrium as
\bea
&&f_{{\bf k},\lambda}(\mathbf{r},t)-  f^0_{k,\lambda}(T(\mathbf{r}))=
 f^0_{k,\lambda}(1-f^0_{k,\lambda})\times  \\ &&\quad\quad\quad\quad\quad\times
{\bf F} \cdot \left[ \hat {\bf v}_{{\bf k},\lambda} g_\parallel(\lambda,\tilde{k}) +\hat {\bf v}_{{\bf k},\lambda}\times {\bf e_z} g_\perp(\lambda,\tk)\right],\nn
\eea
where $g_{\parallel,\perp}$ are dimensionless functions of $\lambda$ and the modulus of $\tilde{\bf k}=v{\bf k}/T$ only, while the dependence on time is understood to be implicit.
The electrical and the heat current associated with such a deviation from equilibrium are given by
\bea
\label{currents1}
\left(\begin{array}{c} \mathbf{J}/e \\ \mathbf{Q}/T
\end{array} \right)
=\frac{4}{2}\frac{T^2}{v} \left(
\begin{array}{c}
{\bf F}\left \langle g_{\parallel} | \phi^E\right \rangle +
{\bf e_z}\times {\bf F} \left \langle g_{\perp} | \phi^E\right \rangle \\
{\bf F} \left \langle g_{\parallel} | \phi^T\right \rangle  +
{\bf e_z}\times {\bf F} \left \langle g_{\perp} | \phi^T\right\rangle
\end{array} \right  ).
\eea
where the factor 4 accounts for both the spin and valley degeneracy, and the inner product
\begin{equation}
\label{innerprod}
\langle g_{1}|g_{2}\rangle \equiv \sum_\lambda \int \frac{d^{2}\tk}{(2\pi )^{2}}%
g_{1}(\tk,\lambda)g_{2}(\tk,\lambda).
\end{equation}%
has been defined.

The linearized Boltzmann equation can be cast in the generic form
\begin{eqnarray}\label{matrixboltz}
&&\left ( \begin{array}{cc} {\cal M} & -\mathcal{B}\\ \mathcal{B} & {\cal M}\end{array}\right) \left ( \begin{array}{c} g_\parallel\\ g_\perp\end{array}\right)
= \left ( \begin{array}{c} \phi^{E,T}\\0\end{array} \right),
\eea
\bea
{\rm where }\quad {\cal M}={\cal C}+{\cal D}-i\Omega\;.
\end{eqnarray}
The linear operators ${\cal C}$ and $\cal D$ are the two components of the collision operator ${\cal I}_{\rm coll}$ describing scattering due to inelastic Coulomb collisions and impurities, respectively.
Explicit expressions have been given in Refs.~\cite{qcgraphene,MHDgraphene}. The dynamical term $\Omega$ acts like
\bea
[\Omega g]({\bf k},\lambda)= \omega\,f^0_{k,\lambda}(1-f^0_{k,\lambda}) g({\bf k},\lambda),
\eea
and arises due to the time derivative in \eqref{Boltzmann}. Finally, ${\cal B}$ describes the deflection due to the magnetic field, and consequently, it is the only term which mixes $g_\parallel$ and $g_\perp$. Typical eigenvalues of these operators correspond to the inelastic and elastic scattering rates $\tau^{-1}_{\rm ee} $, 
$\tau^{-1}_{\rm imp}$, 
the a.c. frequency $\omega$, and the typical cyclotron frequency of thermal particles $\tau^{-1}_B\equiv \omega_c^{\rm typ}\sim v^2 eB/cT$.
One can show~\cite{ziman,yaffe} that the linear operators defined above are Hermitian with respect to the inner product \eqref{innerprod}.

Here we concentrate on the regime of high temperatures, low frequencies and weak fields such that $\tau^{-1}_{\rm ee}$ dominates among the above rates.

In the absence of a magnetic field one easily finds from the above the conductivity, ${\bf J}=\sigma_{xx}{\bf E}$ or the thermopower, ${\bf Q}=\alpha_{xx}{\bf E}$, with
\bea
\label{sigmaxx}
\sigma_{xx}=2\frac{e^2}{\hbar}\left\langle \phi^E|{\cal M}^{-1}|\phi^E\right\rangle,
\eea
\bea
\alpha_{xx}=2\frac{e k_B}{\hbar}\left\langle \phi^E|{\cal M}^{-1}|\phi^T\right\rangle.
\eea

In the presence of a magnetic field this generalizes to ${\bf J}=\sigma_{xx}{\bf E}+\sigma_{xy}({\bf E}\times {\bf e_z})$, with
\bea
\sigma_{xx}=2\frac{e^2}{\hbar}\left\langle \phi^E|({\cal M+BM}^{-1}{\cal B})^{-1}|\phi^E\right\rangle,
\eea
\bea
\sigma_{xy}=2\frac{e^2}{\hbar}\left\langle \phi^E|{\cal M}^{-1}{\cal B}({\cal M+BM}^{-1}{\cal B})^{-1}|\phi^E\right\rangle.
\eea

The above formalism is completely general and does not use any particular properties of graphene. In the sequel we will analyze a remarkable simplification which arises in the hydrodynamic regime in $d=2$. Further, we will show how relativistic hydrodynamics emerges in the case of a massless linear dispersion relation.

\subsection{Divergent forward scattering in 2d}\label{divergence}
It has long been noticed that interactions in $d\leq 2$ lead to singular forward scattering between particles with a linearized spectrum~\cite{wilkins,chubukov}. The physical reason for a diverging cross-section is that collinearly moving particles share the same velocity to the extent that non-linear corrections to the dispersion can be neglected. Thus the interaction time in forward scattering processes diverges, leading to a singular amplitude. While in higher dimensions this effect is irrelevant, being suppressed by its small phase space volume, it emerges as a logarithmic singularity in $d=2$. The singularity is usually cut off by non-linearities in the spectrum, which in graphene are predominantly due to renormalization effects of Coulomb origin. In the case of RPA-screened Coulomb interactions in graphene there is another cut-off mechanism due to the vanishing of the full Coulomb propagator in exactly collinear scattering processes.
Both mechanisms cut off the logarithmic divergence at small scales and lead to an extra factor of $\log(1/\alpha)$ in the typical inelastic relaxation rates, i.e., $\langle g |{\cal C}|g \rangle/\langle g |g \rangle \sim \alpha^2\log(1/\alpha) T$.

The strong forward scattering has an interesting physical consequence: It
leads to an effective equilibration among collinearly moving particles. In
other words the distribution function will tend to a Gibbs distribution
when it is restricted to quasiparticles
with a particular direction of motion,
${\bf \hat{v}}_{{\bf k},\lambda}={\bf e}_\phi$, where ${\bf e}_\phi$ is a
unit vector in the plane. However, the parameters of these one-dimensional quasi-equilibria will in general depend on the angle of motion with respect to the applied field. As always, the equilibrium parameters ($T,\vec{u},\mu$) are conjugate to the conserved quantities, i.e., energy, momentum and charge. However, for massless relativistic particles, there is a further pseudo-conservation law which holds for weak interactions ($\alpha\ll 1$). In this limit, we can restrict ourselves to the Born approximation and evaluate the inelastic collision term with Fermi's Golden rule. Within this approximation, the phase space for two-particle processes ($1+2\to 3+4$) in which the total chirality changes ($\lambda_1+\lambda_2 \neq \lambda_3+\lambda_4$) vanishes exactly~\cite{ssqhe,qcgraphene}. In other words, to leading order in $\alpha^2$, the particle and hole number is separately conserved, which leads to an extra pseudo-equilibrium parameter, $\phi$. We may call it a gravitational potential, since it couples equally to electrons and holes with positive energy.~\footnote{In this paper we only consider perturbations ${\bf E}$ and $\nabla T$ which do not change the overall number of particles and holes, but only redistribute them between different wavevectors. We thus do not have to worry about the very long relaxation time $\tau_{\rm ph}\sim \tau_{\rm ee}/\alpha^2$ which is needed to equilibrate the relative number of particles and holes. The situation is different, however, for scalar perturbations which do alter the total chirality~\cite{viscosity}.}

For a vectorial driving term (an electric field or a thermal gradient) we thus expect
\begin{eqnarray}
f_{{\bf k},\lambda}&\approx&
 \frac{1}{\exp\left[\frac{\varepsilon_{{\bf k},\lambda}-\delta{\bf u}[\hat{\bf v}_{{\bf k},\lambda}]\cdot {\bf k}-(\mu+\delta\mu[\hat{\bf v}_{{\bf k},\lambda}])-\lambda\delta\phi[\hat{\bf v}_{{\bf k},\lambda}]}{T+\delta T[\hat{\bf v}_{{\bf k},\lambda}]}\right]+1}\nn\\
&\approx& f_{k,\lambda}^{0}+
T^{-1}f^0_{k,\lambda}(1-f^0_{k,\lambda})  \\
& \times &\bf{F}\cdot \hat{\bf v}_{{\bf k},\lambda}\left[\lambda k\,\overline{\delta u}+ \overline{\delta \mu}+\lambda \overline{\delta\phi} +(\varepsilon_{{\bf k},\lambda}-\mu)\frac{\overline{\delta T}}{T} \right],\nn
\label{directionaleq}
\end{eqnarray}
where we have taken into account that $\delta T,\delta\mu,\delta u,\delta\phi$ can depend on the driving force ${\bf F}$ only like $\delta T[\hat{\bf v}]={\bf F}\cdot \hat{\bf v}\,\overline{\delta T}$, etc.~\footnote{For a scalar driving force such as $\nabla\cdot {\vec u}$ relevant in the case of bulk viscosity, the prefactor ${\bf F}\cdot \hat{\bf v}$ in (\ref{directionaleq}) is absent, while in the case of a tensorial perturbation such as $\partial u_i/\partial x_j$, relevant for shear viscosity, the prefactor has to be chosen as $\sum_{ij} \hat{v}_i \hat{v}_j\partial u_i/\partial x_j$.}

Note that in the case of a linear dispersion $\varepsilon_{{\bf k},\lambda}=\lambda k v$, the perturbation $\overline{\delta T}$ is redundant since it can be expressed as a combination of two perturbations with $\overline{\delta u}/v=-\overline{\delta\mu}/\mu=\overline{\delta T}/T$. Below we will thus concentrate on the three soft modes~\footnote{In the presence of a perpendicular magnetic field there are additional term analogous to the above, but with $\hat{\bf v} \to {\bf e_z}\times \hat{\bf v}$, and corresponding coefficients $\overline{\delta\mu}_\perp,\overline{\delta\phi}_\perp,\overline{\delta u}_\perp$.}
\bea
\label{g0}
g_0(\lambda,\tk) &=& \lambda \tk,\\
\label{g1}
g_1(\lambda,\tk) &=& 1,\\
\label{g2}
g_2(\lambda,\tk) &=& \lambda.
\eea

To logarithmic accuracy we can restrict the deviations from equilibrium to the above ansatz, since these modes are the only ones with a relaxation rate of order $\alpha^2 T$ (or smaller), while all other modes decay with a rate of order ${\cal O}(\alpha^2 \log(1/\alpha) T)$.

Notice that the response functions $\phi^{E,T}$ \eqref{phis} are simply linear combinations of $g_0$ and $g_1$, multiplied by $f^0_{k,\lambda}(1-f^0_{k,\lambda})$. The basis of soft modes is thus perfectly suited to capture the essentials of thermo-electric response.

In a general two-dimensional Fermi liquid with arbitrary interactions, one may still linearize the spectrum around $E_F$. However, the non-linearity in the dispersion of typical thermal excitations will be of order $T/E_F$, which now cut off the divergence, leading to an extra factor $\log(E_F/T)$ as noticed in Ref.~\cite{wilkins}. However, those authors concluded incorrectly that this logarithm would also show up in the transport relaxation time. Instead, we see here that there are always the above soft modes which decay with a slower rate without logarithmic enhancement, because they represent deviations preserving unidirectional equilibrium which kill the logarithmic divergence in the scattering rate.

\subsection{Recovering relativistic Hydrodynamics}
An entirely different approach to transport in the collision-dominated regime is provided by hydrodynamics. In the present case the relativistic nature of the fluid has to be taken into account, and one should analyze
the equations of motion
\begin{eqnarray}
\p_\beta {J}^{\beta} &=& 0,\\
\label{momentumcons}
\p_\beta {T}^{\beta\alpha}&=& F^{\alpha\gamma}{J}_\gamma +\tau_{\rm imp}^{-1}\delta_{\beta 0}{T}^{\beta\alpha},
\end{eqnarray}
where the stress energy tensor and the current of a relativistic fluid are given by
\begin{eqnarray}
T^{\mu\nu} &=& (\vep+P) u^\mu u^\nu + P g^{\mu\nu} +\tau^{\mu\nu},
\label{e0} \\
J^\mu &=& \rho u^\mu + \nu^\mu,\\
\nu^\mu&=&-\sigma_Q\left[ T(g^{\mu\lambda}+u^\mu
u^\lambda)\partial_\lambda\left(\frac{\mu}{T}\right)-F^{\mu\lambda}u_\lambda\right].
\label{nu-constit}
\end{eqnarray}
Here $u^\mu$ is the three-velocity of the energy density.
We have also included a phenomenological momentum relaxation rate due to impurities, $\tau_{\rm imp}^{-1}$, which explicitly breaks the Lorentz invariance of course. The correctness of this procedure for weak disorder will be confirmed below.

The Reynolds-tensor $\tau^{\mu\nu}$ describing viscous forces can be neglected for the response in the long wavelength limit.
The contribution $\nu^\mu$ to the electrical current corrects for the presence of energy currents that do not carry charge. The form (\ref{nu-constit}) can be inferred by generalizing arguments by Landau and Lifshitz~\cite{LL} for $B=0$, invoking covariance and the second law of thermodynamics~\cite{nernst,cyclotron}, which requires that the divergence of the entropy current be always positive. The argument leading to (\ref{nu-constit}) is valid at small fields $B$, where it is allowed to linearize in $B$.
However, it leaves undetermined the phenomenological conductivity parameter $\sigma_Q>0$.

The hydrodynamic approach allows us to obtain all the low frequency thermoelectric response by linearizing the equations of motions around equilibrium and reading off the response functions following the recipes of Ref.~\cite{km}. Only thermodynamic data and the single coefficient $\sigma_Q$ enter into those expressions, which are discussed in detail in Refs.~\cite{nernst,cyclotron}. This remarkable property of relativistic hydrodynamics, which strongly constrains the form of the response, is fully confirmed by the Boltzmann approach, as we discuss now.

To establish the connection between hydrodynamics and the microscopic analysis, it is central to notice that the "momentum mode" $g_0$ is an exact zero mode of the collision operator ${\cal C}$, ${\cal C}g_0=0$. This is so because $g_0$ describes the equilibrium of a system which moves at speed $\overline{\delta u}$ in the direction of the driving force. Since equilibrium is preserved by the translation invariant Coulomb interactions the perturbation $g_0$ can only decay due to impurity scattering, magnetic fields or the a.c. driving. In the hydrodynamic regime the latter all live on much longer time scales than the relaxation time of any other modes.
It turns out that in this regime it is essentially the dynamics of the momentum mode which governs the magnetoresponse.

The Boltzmann formalism provides an explicit expression for the transport coefficient in the form
\bea
\label{sigmaQ}
\sigma_Q=\sigma_{xx}-\sigma_{\rm mom},
\eea
where, in the absence of impurities,
\bea
\sigma_{\rm mom} =\frac{2e^2}{\hbar} \frac{\left\langle \phi^E| g_0 \right\rangle^2}{\left\langle g_0| \Omega |g_0 \right\rangle}= \frac{e^2}{\hbar}\frac{1}{-i\omega}\frac{\rho^2v^2}{\epsilon+P}
\eea
is the conductivity contribution due to the momentum mode. Here, $\epsilon$ is the energy density and $P$ is the pressure of the free electron gas.
In general $\sigma_{\rm mom}$ diverges in the d.c. limit $\omega\to 0$.
However, at particle-hole symmetry ($\mu=0$) the coupling to the momentum mode vanishes, $\left\langle \phi^E| g_0 \right\rangle=0$, and $\sigma_Q$ is seen to equal the d.c. conductivity in the absence of impurities. The interesting fact that this is in fact finite will be discussed below. To logarithmic accuracy one finds the result~\cite{qcgraphene}
\bea
\label{sigmauniv}
\sigma_Q(\mu=0,\omega=0)= \frac{2e^2}{\hbar} \frac{\left\langle \phi^E| g_1 \right\rangle^2}{\left\langle g_1| {\cal C} |g_1 \right\rangle} = \frac{0.121}{\alpha^2}\frac{e^2}{\hbar}.
\eea
For general $\mu\neq 0$ one can obtain a logarithmically accurate approximation for $\sigma_{xx}$ and $\sigma_Q$ by observing that $\sigma_{xx}$ is the maximum of the functional
\bea
Q\left[g\right]=\frac{2 e^2}{\hbar} \left[ 2\left\langle \phi^E| g \right\rangle -\left\langle g| {\cal C} |g \right\rangle \right],
\eea
and extremizing it over the linear space spanned by the three soft modes $g_{0},g_1,g_2$, Eqs.~(\ref{g0}-\ref{g2}) discussed above.
In full generality, one finds that $\sigma_Q$ is a scaling function of $\mu/T$ which decays as a power law for $\mu\gg T$~\cite{MHDgraphene}.

The response functions $\sigma_{xx}$, $\alpha_{xx}$ and the thermal conductivity $\kappa_{xx}$ have all the same structure in the absence of a magnetic field. There is a contribution from the momentum mode, which generally diverges in the clean d.c. limit, and in addition there is an extra piece proportional to $\sigma_Q$ which is contributed by the Coulomb relaxing modes. An explicit calculation~\cite{MHDgraphene} shows that the relative deviations between hydrodynamics and the microscopic Boltzmann calculation are of higher order in ${\cal O}(\omega\tau_{\rm ee},\tau_{\rm ee}/\tau_{\rm imp})$, which are indeed small in the hydrodynamic regime. The same statement holds also for the case with magnetic field, where relative corrections are suppressed by factors of ${\cal O}(\tau_{\rm ee}/\tau_B)$.

\subsection{Impurity scattering}
The momentum relaxation rate which was added in the hydrodynamic formulation (\ref{momentumcons}) in an ad hoc manner, turns out to reproduce the leading disorder effects of the Boltzmann theory
provided that the relaxation rate is taken to be the inverse of the elastic lifetime of the momentum mode,
\bea \label{tauimp}
\tau_{\rm imp}^{-1} =  \frac{ \left\langle g_0|{\cal D}|g_0\right\rangle}{\left\langle g_0|\Omega/\omega|g_0\right\rangle} \propto \frac{e^4}{\kappa^2} \frac{n_{\rm imp}}{\max[T,|\mu|]}.
\eea
The final estimate applies to charged, unscreened impurities of density $n_{\rm imp}$. The lowest order corrections to the clean hydrodynamic results are simply obtained by replacing $\omega\to \omega +i\tau_{\rm imp}^{-1}$. In particular the contribution of the momentum mode to the d.c. conductivity takes the generalized Drude form
\bea
\label{sigmamom}
\sigma_{\rm mom} = \frac{e^2}{\hbar} \tau_{\rm imp} \frac{\rho^2v^2}{\epsilon+P},
\eea
where the band mass $m$ is replaced by $({\epsilon+P})/(\rho v^2)$.

\subsection{Crossover in the electrical conductivity}

The above described approximation of restricting to the soft modes lends itself naturally to the description of an experimentally relevant situation~\cite{Fuhrer}: the crossover observed when the system is driven from the critical, relativistic regime ($\mu\ll T$) to the conventional Fermi liquid regime ($\mu\gg T$) by adjusting the doping level, see Refs.~\cite{joerg,MHDgraphene}. In the Fermi liquid regime we can make connection with existing results~\cite{Nomura,DasSarmaGalitski}.
It is worth noticing that the asymptotics for $\mu \ll T$ and $\mu \gg T$ is already well described by only two modes, $g_0$ and $g_1$. Indeed, the "chirality mode" $g_2$ either contributes very little to the current  since it is hardly excited by the electrical field  at $\mu \ll T$, or it is essentially identical to the "chemical potential mode" $g_1$ when $\mu \gg T$, where essentially only one type of charge carriers with a definite sign of $\lambda$ is present.
The crossover is thus expected to be reasonably well captured by this two-mode approximation, which was carried out in detail in Ref.~\cite{MHDgraphene}

For low doping, in the critical regime $\mu \ll T$, the inelastic scattering time behaves as
\bea
\label{tauee}
\tau_{ee}^{-1}=\alpha^2 T
\eea
reflecting the quantum criticality of the system~\cite{joerg,qcgraphene}. Upon doping, the Coulomb scattering rate decreases and ultimately behaves as $\tau_{ee}^{-1}\propto \alpha^2 \frac{T^2}{\mu}$ for $\mu \gg T$ if screening is taken into account. This is the standard inelastic scattering rate in a Fermi liquid.
The relaxation due to impurities behaves very differently as a function of $T$ and $\mu$, as seen from Eq.~\eqref{tauimp}. The increase of the scattering rate with decreasing temperature reflects the increased crosssection for Coulomb scattering at low energies. The crossover from a collision-dominated, critical conductivity where $\sigma\approx \sigma_Q(\mu=0)\sim e^2/\hbar\alpha^2$ to an impurity limited conductivity which is dominated by the contribution from the momentum mode (\ref{sigmamom}) occurs when the density  of thermally excited carriers, $\rho_{\rm th}\sim T^2/(\hbar v)^2$, (holes and electrons) equals the density of impurities, $n_{\rm imp}$.

Of course it is possible to go beyond the logarithmic approximation based on two or three modes, if the logarithmic divergence is properly regularized. In particular, it was shown in Ref.~\cite{MHDgraphene} that in the weak disorder limit one can absorb the effect of all modes other than $g_0, g_1$ into one (or two in the case of strong magnetic fields) effective frequency-dependent matrix element(s) entering all the response functions. This is a more precise version of the hydrodynamic statement that at low fields the whole response contains solely $\sigma_Q$ as non-thermodynamic parameter.

\subsection{Criticality at the Dirac point}
The fact that $\sigma_{xx}$ is finite in a clean, undoped system ($\mu=0$) is a peculiar feature which is also known from neutral plasmas. It is possible because at particle-hole symmetry, an electric field does not excite a momentum flow, which would not decay under Coulomb collisions and thus yield an infinite conductivity. Instead, the electron and hole currents decay within the finite relaxation time $\tau_{\rm ee}$ due to their mutual friction. Since there is no dimensionless parameter in the problem except for $\alpha$, the universal collision-dominated scaling \eqref{sigmauniv}, $\sigma\sim e^2/h\alpha^2$, can be predicted simply from dimensional considerations~\cite{Ryzhii}.

It is also interesting to study other thermoelectric response functions at the Dirac point and their crossover to to Fermi liquid behavior at large doping. For example one finds an anomalously large Mott ratio, $-\alpha_{xx}/(d\sigma/d\mu)$, in the regime $\mu\ll T$, as well as strong modifications of the Wiedemann-Frantz law, given that the thermal conductivity diverges in a clean system as $\mu\to 0$. In both cases the deviation from Fermi liquid behavior at low doping, is diminishing at $T\sim \mu$, recovering the usual Fermi liquid characteristics at high doping levels.

Equally unusual behavior of clean undoped graphene can be expected with respect to the shear viscosity, a quantity which is of great interest recently in the context of another relativistic fluid: the quark gluon plasma where the shear viscosity is found to be anomalously small, approaching a lower bound conjectured from a general holographic correspondence with black hole problems and their area law. The large critical scaling of $\tau^{-1}_{\rm ee}\sim T$ in graphene also suggests an exceptionally small value of the shear viscosity at the Dirac point (since $\eta\sim \tau_{\rm ee}$, as opposed to the Fermi liquid regime $\mu\gg T$. This is indeed confirmed by explicit calculations.~\cite{viscosity}

\subsection{Magnetotransport and cyclotron resonance}
An interesting prediction of relativistic hydrodynamics is the appearance of a collective cyclotron resonance~\cite{cyclotron}, whose existence is confirmed by the Boltzmann approach. As we have mentioned before, the dynamics of the momentum mode dominates most of the thermo-electric response, since it has the longest life time. In the presence of magnetic fields, impurities and a finite a.c. frequency, $g_0$ is not a zero mode of (\ref{matrixboltz}) any more, but there is still a unique pair of modes corresponding to the smallest eigenvalue of that operator (the original pair of zero modes $g_\parallel=g_\perp=g_0$ is split into two modes).

Within first order perturbation theory, projecting the operators in Eq.~\eqref{matrixboltz} to the momentum mode, one expects a resonance in the clean material at
\bea
\label{omegac}
\omega_c = \pm \frac{\left\langle g_0|{\cal B}|g_0\right\rangle}
{\left\langle g_0|(\Omega/\omega)|g_0\right\rangle}=\pm \frac{e B \rho v^2}{c(\epsilon+P)}\,.
\eea
Notice the similarity to the expression $\omega_c=e B/c m$ for a standard cyclotron resonance.
Again the band mass is replaced by $m\to(\epsilon+P)/\rho v^2$, and $\omega_c$ is proportional to the density, which reflects the collective nature of this hydrodynamic mode.

Including weak impurities to first order perturbation theory, and the coupling of the momentum mode to the faster relaxing modes in second order in $B$, one finds an imaginary part to the low eigenvalue. The frequency pair for which the operator on the left hand side of \eqref{matrixboltz} is singular, becomes
\bea
\label{omegapole}
\omega_{\rm pole}\approx \pm \omega_c-i(\gamma+\tau_{\rm imp}^{-1}),
\eea
where the collision induced damping is given by
\bea
\label{gamma}
\gamma= \frac{\sigma_QB^2v^2}{c^2(\epsilon+P)}.
\eea
The pole $\omega_{\rm pole}$ shows up in all frequency-dependent response functions as a resonance $\sim (\omega-\omega_{\rm pole})^{-1}$.
The fact that $\gamma$ has such a simple expression in terms of the coefficient $\sigma_Q$ defined above in \eqref{sigmaQ}, is a special feature of the massless Dirac spectrum which ensures that ${\cal B}g_0\propto \phi^E$. At $\mu=0$ the above relation (\ref{gamma}) is indeed easy to check in second order perturbation theory, using that $\sigma_Q=\sigma$ is given by (\ref{sigmaxx}). A general proof of (\ref{gamma}) was given in Ref.~\cite{MHDgraphene}. It is also shown there that the property ${\cal B}g_0\propto \phi^E$ is ultimately responsible for the fact that the relativistic hydrodynamics emerges in the microscopic theory.

The interaction induced broadening $\gamma$ is already predicted by hydrodynamics~\cite{cyclotron}, and is exactly reproduced by the Boltzmann formalism. However, the latter allows one also to go beyond small magnetic fields, an assumption that had to be made when deriving the relation \eqref{nu-constit}. While there is still a collective cyclotron resonance due to the damped dynamics of the original momentum mode, the pole must be found from the solution of
\bea
{\rm Det} \left ( \begin{array}{cc} {\cal M}(\omega_{\rm pole}) & -\mathcal{B}\\ \mathcal{B} & {\cal M}(\omega_{\rm pole})\end{array}\right)=0.
\eea
In perturbation theory one can establish that $\Re[\omega_{\rm pole}]$ exceeds the lowest order result (\ref{omegapole}), while (\ref{gamma}) overestimates the damping $\Im[\omega_{\rm pole}]$.
A numerical solution of the above secular equation shows that at high fields, where $\tau^{-1}_B\sim \tau^{-1}_{\rm ee}$ this trend persists.~\cite{MHDgraphene}

Interestingly, this behavior is qualitatively similar as in nearly critical, relativistic conformal field gauge theories which are exactly solvable thanks to the AdS-CFT correspondence.~\cite{AdsCFT}

Clearly, the magnetic fields need to be such that the Landau level spacing $E_{\rm LL}=\hbar v\sqrt{2 e B/\hbar c}$ is still small compared to the temperature. Otherwise, interference effects and the Quantum Hall effect start to set in, which cannot be described on the level of a semiclassical Boltzmann theory. However, at small $\alpha$, this still leaves a large window $\alpha^2 T<\tau^{-1}_B<T$ for the Boltzmann theory to apply in a regime where the fields have to be considered large and the straightforward hydrodynamic description with the constitutive relation (\ref{nu-constit}) breaks down.

In conclusion, we have discussed the transport properties of a weakly coupled fluid of massless Dirac particles. The critical regime of low doping and high temperatures is predicted to exhibit several unusual characteristics, such as a universal collision-dominated conductivity, an anomalously low viscosity and a collective, interaction-damped cyclotron mode. We hope that these effects will soon be within experimental reach in very clean, suspended graphene as studied in Refs.~\cite{andrei,stormer}.

The fact that in two dimensions forward scattering is anomalously strong allows one to solve the Boltzmann equation in a simple, physically transparent way which is asymptotically exact for weakly coupled graphene, and is well adapted to describe the crossover to the Fermi liquid regime.

\begin{theacknowledgments}
This research was supported by the Swiss National
Fund for Scientific Research under grant PA002-113151 (MM); by the Deutsche
Forschungsgemeinschaft under grant FR 2627/1-1 (LF);  by the NSF under
grant DMR-0757145 (SS), and by the Ames Laboratory, operated for the U.S.
Department of Energy by Iowa State University under Contract No.
DE-AC02-07CH11358 (JS).
\end{theacknowledgments}

\bibliographystyle{aipproc}   

\end{document}